\documentclass[aps,preprintnumbers,twocolumn,amsmath,nofootinbib,amssymb,floatfix]{revtex4}
\usepackage{graphicx,dcolumn,booktabs,bm}
\usepackage{epstopdf}
\usepackage{longtable,lscape}
\usepackage{txfonts}
\usepackage{overpic}
\usepackage{float}
\usepackage{amssymb}
\usepackage{amsmath}
\usepackage{indentfirst}
\usepackage{feynmf}  
\usepackage{slashed}  
\usepackage{cases}
\usepackage{ulem}
\usepackage{color}
\usepackage{float}
\usepackage{multirow}
\usepackage{tikz}
\usepackage{epsfig,dsfont,amsfonts,amsbsy,mathrsfs}
\usepackage{multirow}
\usepackage{graphicx}  
\usepackage{float}  
\usepackage{subfigure}  
\usepackage{array}
\usepackage{microtype}
\usepackage{pdfpages} 
\usepackage{flushend}
\usepackage[colorlinks, citecolor=blue,anchorcolor=red,menucolor=red, linkcolor=red,filecolor=red,runcolor=red,urlcolor=blue,frenchlinks=true]{hyperref}

\usepackage{braket}

\allowdisplaybreaks[4]

\begin{document}

\title{Probing the nature of $D_1 K$ and $D_2 K$ molecules through $D_s^{(*)}\pi\pi$ and $D_{s0(s1)}\pi$ decays}

\author{Xing-Meng Zhao}
\author{Liu-Lin~Wang}\email{liulinwang@tju.edu.cn}
\author{Ying-Bo He}
\author{Xiao-Yun Wang}
\author{Xiao-Hai Liu}\email{xiaohai.liu@tju.edu.cn}

\affiliation{Center for Joint Quantum Studies and Department of Physics, School of Science, Tianjin University, Tianjin 300350, China}

\date{\today}

\begin{abstract}
We study the two- and three-body decays of the $I=0$ $D_1K$ and $D_2K$ molecular states $T_{c\bar{s}1}^*$ and $T_{c\bar{s}2}^*$ into $D_s^{(*)}$ mesons and pions. Triangle singularities produce narrow peaks in the $D_s^*\pi$ and $D_s\pi$ invariant mass spectra near the $D^*K$ and $DK$ thresholds. The isospin-violating two-body decays $T_{c\bar{s}1}^*\to D_{s1}(2460)\pi^0$, $T_{c\bar{s}2}^*\to D_{s1}(2460)\pi^0$, and $T_{c\bar{s}2}^*\to D_{s0}^*(2317)\pi^0$ exhibit large partial widths, reflecting the strong couplings inherent to molecular states. These predictions, obtained within heavy hadron chiral perturbation theory and the chiral unitary approach, provide complementary signatures for identifying $D_1K$ and $D_2K$ molecules at experiments.
\end{abstract}

\maketitle

\section{Introduction}
Hadron spectroscopy has experienced a renaissance in recent years with the discovery of numerous exotic states that challenge conventional quark model classifications. Many of these so-called $XYZ$ particles are observed near two-hadron thresholds, suggesting possible interpretations as hadronic molecules~\cite{Guo:2017jvc,Olsen:2017bmm,Brambilla:2019esw,Chen:2022asf,Chen:2016qju,Chen:2016spr,Esposito:2016noz}. However, alternative explanations involving kinematic effects, particularly triangle singularities (TS), have gained traction as they can produce resonance-like structures without invoking genuine resonances~\cite{Guo:2019twa}. The TS mechanism, first discussed in the 1960s~\cite{Eden:1966dnq,Landau:1959fi}, occurs when three internal propagators in a triangle Feynman diagram become simultaneously on-shell, leading to characteristic peaks in invariant mass distributions.

This mechanism has been successfully applied to a variety of exotic phenomena~\cite{Liu:2019dqc,Liu:2020orv,Liu:2016xly,Szczepaniak:2015eza,Guo:2015umn,Liu:2015taa,Liu:2015fea,Guo:2016bkl,Bayar:2016ftu,Roca:2017bvy,Liu:2016dli,Liu:2015cah,Cao:2017lui,Dong:2020hxe,Nakamura:2021qvy,Wang:2025zbv}. Notable examples include the large isospin violation in $\eta(1405)\to 3\pi$~\cite{Wu:2011yx,Aceti:2012dj,Achasov:2015uua}, the production of $a_1(1420)$~\cite{COMPASS:2020yhb,Ketzer:2015tqa,Aceti:2016yeb}, the appearance of $Z_c(3900)$ and $Z_c(4020)$ near the $\bar{D}D^*$ and $\bar{D}^*D^*$ thresholds~\cite{Wang:2013cya,Liu:2013vfa,Liu:2014spa,Guo:2014iya}, and so on. For a comprehensive review, we refer the reader to Ref.~\cite{Guo:2019twa}.

The TS mechanism plays a dual role in hadron physics: on one hand, it offers an alternative explanation for near-threshold enhancements that might otherwise be mistaken for new resonances; on the other hand, it provides a tool to probe the internal structure of hadronic molecules through their characteristic kinematic signatures.

A well-established example of molecular states in the charm-strange sector is provided by the $D_{s0}^*(2317)$ and $D_{s1}(2460)$ mesons. Discovered by the BaBar and CLEO collaborations in the early 2000s, these states have masses lying about 40–50 MeV below the $DK$ and $D^*K$ thresholds, respectively, which is in sharp contrast to the predictions of naive quark models~\cite{Aubert:2003fg,Besson:2003cp,Godfrey:1985xj}. Over the past two decades, a growing body of evidence from chiral unitary approach, lattice QCD, and effective Lagrangian analyses has converged to support their interpretation as hadronic molecules composed of $DK$ and $D^*K$, with spin-parities $J^P = 0^+$ and $1^+$, respectively~\cite{Liu:2012zya,Mohler:2013rwa,Lang:2014yfa,Torres:2014vna,Lang:2015hza,Moir:2016srx,Bali:2017pdv,Albaladejo:2016lbb,Du:2017zvv,Albaladejo:2018mhb,Guo:2018kno,Barnes:2003dj,vanBeveren:2003kd,Kolomeitsev:2003ac,Chen:2004dy,Guo:2006fu,Guo:2006rp,Yang:2021tvc}. In this picture, the $D^{(*)}K$ interaction is strong enough to form a bound state, and the presence of $u(d)$ quarks in the constituents gives rise to direct strong isospin-violating decay modes $D_{s0}^* \to D_s \pi^0$ and $D_{s1} \to D_s^* \pi^0$~\cite{Faessler:2007us,Fu:2021wde,Lutz:2007sk, Guo:2008gp,Liu:2012zya,Cleven:2014oka,Guo:2018kno}. The $D_{s0}^*(2317)$ and $D_{s1}(2460)$ thus serve as prototypical examples of $DK$ and $D^*K$ hadronic molecules in nature.

The $D_1(2420)K$ and $D_2(2460)K$ molecular states that are the focus of this work are intimately connected to the $D_{s0}^*(2317)$ and $D_{s1}(2460)$ through the chiral and heavy-quark symmetry that underlies the dynamics of heavy-light mesons interacting with Nambu–Goldstone bosons (NGBs). Within the framework of unitary chiral perturbation theory (UChPT), the same coupled-channel interactions that dynamically generate the $DK$ and $D^*K$ bound states also produce poles in the $D_1K$ and $D_2K$ systems, with $D_1$ and $D_2$ being the $P$-wave charm mesons (with $j_l = 3/2$ in the heavy-quark limit)~\cite{Wang:2025jcq}. The $D_1K$ and $D_2K$ states can therefore be viewed as the natural heavier partners of the $D_{s0}^*(2317)$ and $D_{s1}(2460)$, involving $P$-wave charm mesons instead of $S$-wave ones, while sharing the same underlying mechanism of molecular binding. 
In this work, we focus on the $D_1 K$ and $D_2 K$ molecular states with isospin $I=0$, denoted as $T_{c\bar{s}1}^*$ and $T_{c\bar{s}2}^*$, respectively. In our previous work~\cite{Wang:2025jcq}, poles near 2860 MeV have been identified as $D_1 K$ and $D_2 K$ molecular states, which may correspond to the experimentally observed $D_{sJ}(2860)$~\cite{BaBar:2006gme,LHCb:2014ott}.

A useful way to scrutinize the internal structure of these exotic states is to analyze decay channels that are highly sensitive to long-range hadron–hadron dynamics. Three-body decays proceeding via hadronic loops can produce kinematic TS that imprint narrow enhancements in invariant-mass spectra. Within the molecular framework, $T_{c\bar{s}1}^*$ is predicted to decay into $D_s^*\pi\pi$, while $T_{c\bar{s}2}^*$ can decay into both $D_s^*\pi\pi$ and $D_s\pi\pi$ through rescattering mechanisms that are naturally embedded in the loop amplitudes.

In addition to three-body modes, isospin-violating two-body decays provide complementary probes. Although these transitions are nominally suppressed, they can be significantly enhanced when the parent state lies close to the relevant hadronic threshold and couples strongly to its constituents. A representative example is $T_{c\bar{s}1}^{*}\to D_{s1}(2460)\pi^0$, which proceeds via two interfering diagrams whose net interference is destructive. The combined pattern across three-body and isospin-violating two-body channels thus encodes both the long-range molecular dynamics and the impact of kinematic singularities, enabling data-driven discrimination against alternative compact interpretations.

The paper is organized as follows. Section II introduces the effective Lagrangians, coupling assignments, and the construction of the decay amplitudes. Section III presents numerical results, including invariant-mass spectra and decay width estimates for both three-body and isospin-violating two-body channels. Section IV summarizes our conclusions.

\section{Theoretical Framework}

\subsection{Three-body decays of $T_{c\bar{s}1}^*$ and $T_{c\bar{s}2}^*$}

\begin{figure}[htbp]
\centering
\includegraphics[width=1.0\hsize]{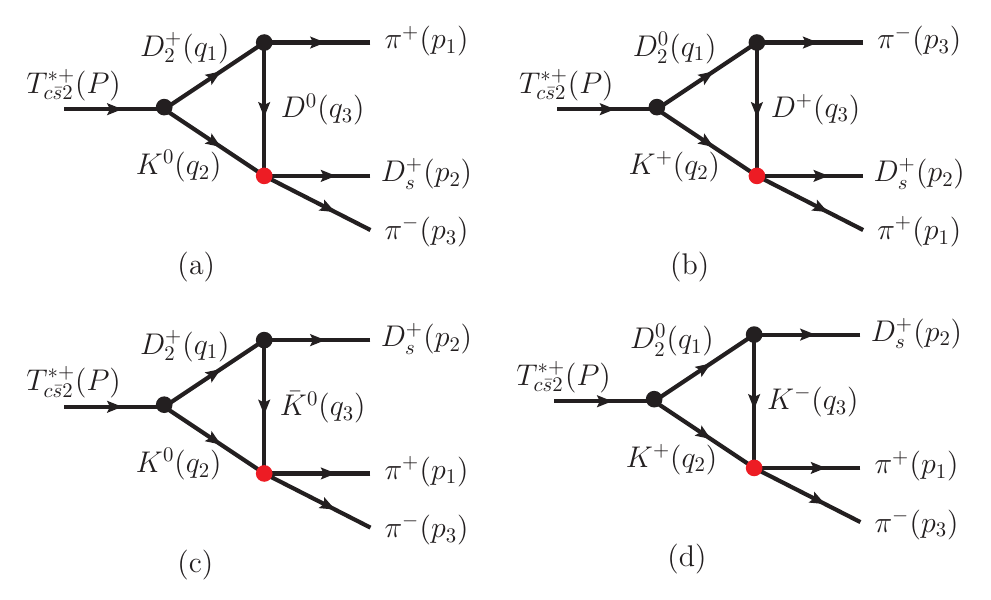}
\caption{Triangle diagrams for $T_{c\bar{s}2}^{*+} \to D_s^{+} \pi^+ \pi^-$: 
(a) $D_2^+ K^0 D^{0}$ loop; (b) $D_2^0 K^+ D^{+}$ loop;
(c) $D_2^+ K^0 \bar{K}^0$ loop; 
(d) $D_2^0 K^+K^-$ loop.}
\label{fig:triangle3}
\end{figure}

We assume $T_{c\bar{s}1}^*$ and $T_{c\bar{s}2}^*$ to be weakly bound $D_1 K$ and $D_2 K$ molecules with isospin $I=0$. 
Their couplings to the constituent mesons can be described by effective Lagrangians
\begin{align}
\mathcal{L}_{T_{c\bar{s}1}^* D_1 K} &= \frac{g_{T_1}} {\sqrt{2}}\, \bar{T}_{c\bar{s}1}^{*\mu} \left( D_{1\mu} ^+ K^0 + D_{1\mu} ^0 K^+ \right) + \text{H.c.}, \label{Eq:lagT1} \\ 
\mathcal{L}_{T_{c\bar{s}2}^* D_2 K} &= \frac{g_{T_2}} {\sqrt{2}}\, \bar{T}_{c\bar{s}2}^{*\mu\nu} \left( D_{2\mu\nu} ^+ K^0 + D_{2\mu\nu} ^0 K^+ \right) + \text{H.c.}, \label{Eq:lagT2}
\end{align}
where $g_{T_1}$ and $g_{T_2}$ denote the corresponding coupling constants. In this work, these couplings are taken from our previous study \cite{Wang:2025jcq}, where they were extracted from the residues at the poles of the scattering amplitudes in the $D_1 K$ and $D_2 K$ channels.
Within the hadronic molecule framework, the $T_{c\bar{s}1}^*$ state is predicted to decay to $D_s^* \pi \pi$, while its counterpart, $T_{c\bar{s}2}^*$, is expected to decay to either $D_s^* \pi \pi$ or $D_s \pi \pi$. The rescattering diagrams that illustrate these three-body decays are shown in Figs.~\ref{fig:triangle1},~\ref{fig:triangle2}, and~\ref{fig:triangle3}.

\subsubsection{$T_{c\bar{s}1}^{*+} \to D_s^{*+} \pi^+ \pi^-$}

\begin{figure}[htbp]
\centering
\includegraphics[width=1.0\hsize]{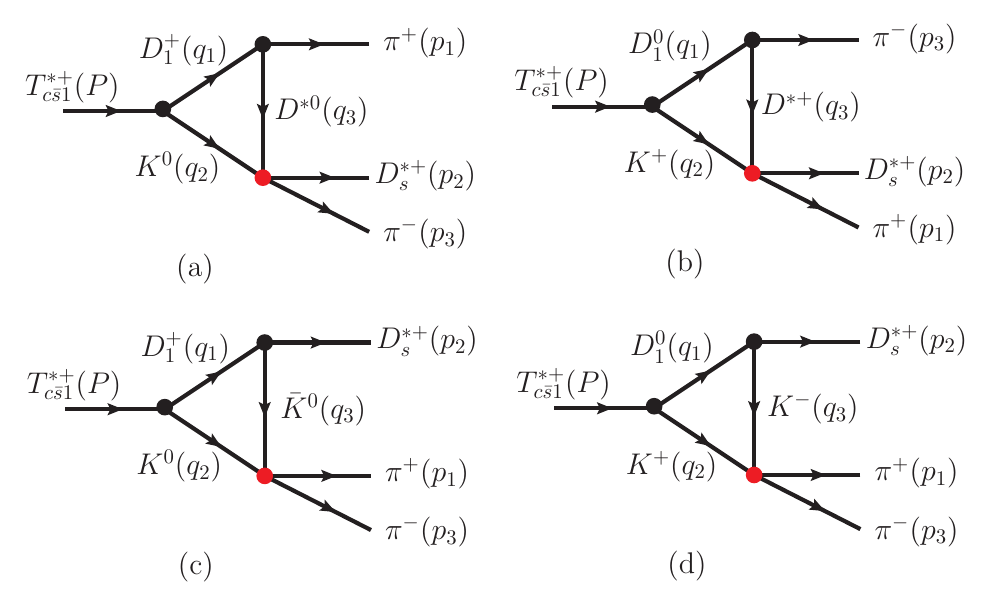}
\caption{Triangle diagrams for $T_{c\bar{s}1}^{*+} \to D_s^{*+} \pi^+ \pi^-$: 
(a) $D_1^+ K^0 D^{*0}$ loop; (b) $D_1^0 K^+ D^{*+}$ loop;
(c) $D_1^+ K^0 \bar{K}^0$ loop; 
(d) $D_1^0 K^+K^-$ loop.}
\label{fig:triangle1}
\end{figure}

The rescattering amplitude of $T_{c\bar{s}1}^{*+} \to D_s^{*+} \pi^+ \pi^-$ via the $D_1^+(q_1) K^0(q_2) D^{*0}(q_3)$-loop in Fig.~\ref{fig:triangle1}(a) is given by 
\begin{align}
&\mathcal{A}^{[D_1^+ K^0 D^{*0}]}_{T_{c\bar{s}1}^{*+}  \to D_s^{*+} \pi^+ \pi^-}=-i\int\frac{d^4 q_1}{(2\pi)^4}\,\frac{\mathcal{A}(T_{c\bar{s}1}^{*+} \to D_1^+ K^0)}{(q_1^2 - m_{D_1}^2 + im_{D_1}\Gamma_{D_1})}
\nonumber \\[2mm]
& \times
\frac{\mathcal{A}(D_1^+ \to D^{*0} \pi^+)}{(q_2^2 - m_{K}^2)}
\,
\frac{\mathcal{A}(D^{*0} K^0 \to D_s^{*+} \pi^-)}{(q_3^2 - m_{D^*}^2)} \mathbb{F}(q_3^2)
, \label{Eq:loopT1}
\end{align}
where the sum over polarizations of the intermediate state is implicit.
In Eq.~(\ref{Eq:loopT1}), a monopole form factor $\mathbb{F}(q_3^2)=(m_{D^*}^2-\Lambda^2)/(q_3^2-\Lambda^2)$ is introduced to account for the off-shell effects and kill the ultraviolet divergence that may appear in the loop integral.
For the intermediate spin-1 state, the sum over polarization takes the form $\sum_{pol.} \epsilon_\mu \epsilon_\nu^*=-g_{\mu\nu}+v_\mu v_\nu$, and we set the velocity of the heavy field $v=(1,\boldsymbol{0})$ for a non-relativistic approximation.  The Breit-Wigner type propagator is introduced in Eq.~(\ref{Eq:loopT1}) to account for the width effect of the intermediate state. In the numerical calculation, we adopt the physical value $\Gamma_{D_1}=31.3$~MeV from the Particle Data Group~\cite{ParticleDataGroup:2024cfk}.

According to Eq.~(\ref{Eq:lagT1}), the vertex function $\mathcal{A}(T_{c\bar{s}1}^{*+} \to D_1^+ K^0)$ in Eq.~(\ref{Eq:loopT1}) reads
\begin{equation}
\mathcal{A}(T_{c\bar{s}1}^{*+} \to D_1^+ K^0) =\frac{g_{T_1}} {\sqrt{2}} \varepsilon(P)\cdot \varepsilon^*(q_1),
\end{equation}
with $\varepsilon(P)$ and $\varepsilon^*(q_1)$ the polarization vectors of $T_{c\bar{s}1}^{*+}$ and $D_1^+$, respectively.

\begin{figure}[htbp]
\centering
\includegraphics[width=1.0\hsize]{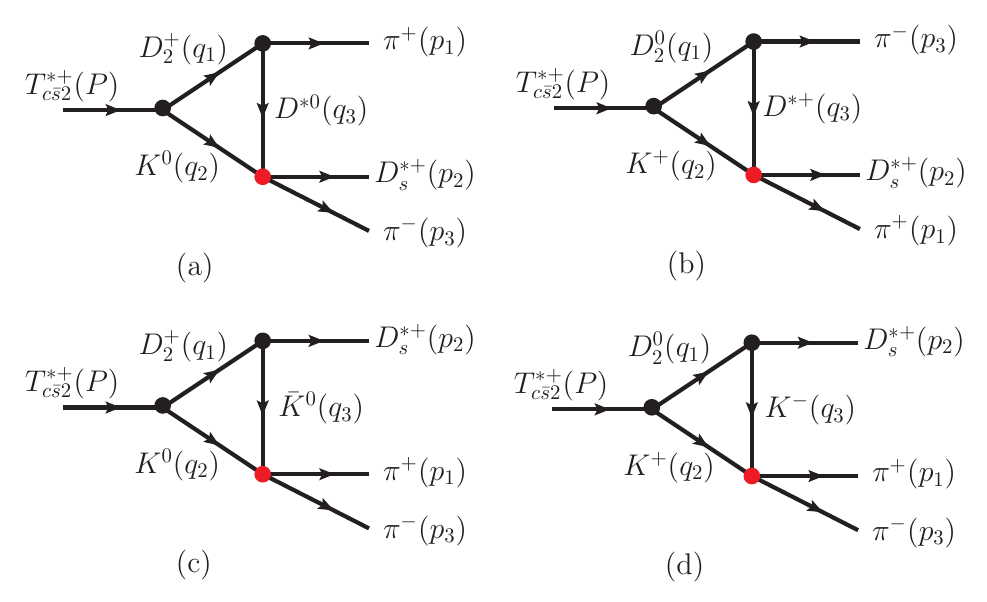}
\caption{Triangle diagrams for $T_{c\bar{s}2}^{*+} \to D_s^{*+} \pi^+ \pi^-$: 
(a) $D_2^+ K^0 D^{*0}$ loop; (b) $D_2^0 K^+ D^{*+}$ loop;
(c) $D_2^+ K^0 \bar{K}^0$ loop; 
(d) $D_2^0 K^+K^-$ loop.}
\label{fig:triangle2}
\end{figure}

The vertex $D_1^+ \to D^{*0} \pi^+$ is a $D$-wave decay described by heavy hadron chiral perturbation theory (HHChPT)~\cite{Casalbuoni:1996pg,Wise:1992hn,Burdman:1992gh,Yan:1992gz}. The relevant strong interaction terms in HHChPT are given by the Lagrangian
\begin{eqnarray}
\label{eq:Lag-LT}
{\cal L}_{\mathcal{T}} &=& i\frac{h^\prime}{\Lambda_\chi} \langle {\bar H}_{a} T_{b}^{\mu} {\gamma}_\nu {\gamma}_5 (D_\mu {\cal A}_\nu + D_\nu {\cal A}_\mu)_{ba} \rangle +\text{H.c.},
\end{eqnarray}
where the heavy superfields are defined as
\begin{equation}
\label{heavyfield}
\begin{aligned}
H_a &= \frac{1+\slashed{v}}{2}\left(P^{\ast\mu}_{a}\gamma_{\mu}-P_a\gamma_5\right), \\[4pt]
T_a^\mu &= \frac{1+\slashed{v}}{2} \Bigg\{ P_{2a}^{\mu\nu}\gamma_\nu
- \sqrt{\frac{3}{2}}\, P_{1a\nu} \gamma_5 \left[g^{\mu\nu} - \frac{1}{3}\gamma^\nu(\gamma^\mu - v^\mu)\right]
\Bigg\}, \\[4pt]
\bar{H}_a &= \gamma_0 H^{\dagger}_a\gamma_0,\qquad 
\bar{T}_a^\mu = \gamma_0 T_a^{\mu\dagger} \gamma_0.
\end{aligned}
\end{equation}
Here, $v$ denotes the heavy meson velocity and $\langle \cdots \rangle$ represents the trace over Dirac matrices. The flavor index $a$ labels light quarks. The $H_a$ doublet contains $P = (D^0, D^+, D_s^+)$ and $P^{\ast} = (D^{\ast 0}, D^{\ast +}, D_s^{\ast +})$ with $s_\ell^p = \frac{1}{2}^-$, where $s_\ell^p$ denotes the total angular momentum and parity of the light degrees of freedom in the heavy-quark limit. For the $s_\ell^p = \frac{3}{2}^+$ $P$-wave heavy-light mesons, the $T_a$ doublet contains $P_{1} = (D_1^0, D_1^+, D_{s1}^+)$ and $P_{2} = (D_2^{0}, D_2^{+}, D_{s2}^{+})$. In Eq.~(\ref{eq:Lag-LT}), ${\cal A}^\mu$ is the chiral axial vector containing the NGBs:
\begin{eqnarray}
{\cal A}_\mu = \frac{1}{2}\left(\xi^\dagger \partial_\mu \xi - \xi \partial_\mu \xi^\dagger\right),
\end{eqnarray}
with
\begin{eqnarray}
\xi &=& e^{i{\cal M}/f_{\pi}},\\[4pt]
{\cal M} &=& \left( \begin{array}{ccc}
\frac{1}{\sqrt{2}}\pi^0 + \frac{1}{\sqrt{6}}\eta & \pi^+ & K^+ \\
\pi^- & -\frac{1}{\sqrt{2}}\pi^0 + \frac{1}{\sqrt{6}}\eta & K^0 \\
K^- & \bar{K}^0 & -\sqrt{\frac{2}{3}}\eta
\end{array} \right).
\end{eqnarray}

The vertex function for $D_1^+ \to D^{*0} \pi^+$ is then given by
\begin{align}
&\mathcal{A}(D_1^+ \to D^{*0} \pi^+)
= \frac{4}{\sqrt{6}} \frac{h^\prime}{\Lambda_\chi  f_\pi} \sqrt{m_{D_1} m_{D^*}} \, 
 \nonumber\\
&\times \Big\{
3\, p_1^\lambda p_1^\nu+ \big[ (v \cdot p_1)^2 - p_1^2 \big] g^{\lambda \nu} 
\Big\} 
\, \varepsilon_\nu(q_1)\, \varepsilon_\lambda^*(q_3),
\end{align}
with $\varepsilon^*(q_3)$ the polarization vector of $D^{*0}$, the coupling constant $h^\prime \simeq 0.43$, $\Lambda_\chi=1$ GeV, and $f_\pi=132$ MeV~\cite{Colangelo:2012xi,Colangelo:2005gb}.

For the vertices $D^{(*)} K \to D_s^{(*)} \pi$ depicted in Figs.~\ref{fig:triangle1} and ~\ref{fig:triangle2}, we employ amplitudes derived from the UChPT~\cite{Oller:2000fj,Oller:2000ma,Oller:1997ng}. In this work, we focus specifically on the $S$-wave component of the $D^{(*)} K$-$D_s^{(*)} \pi$ coupled-channel scattering. The amplitude $\mathcal{A}(D^{(*)} K \to D_s^{(*)} \pi)$ is then obtained as
\begin{equation}
   \mathcal{A}(D^{(*)} K \to D_s^{(*)} \pi) =[(1 - V G)^{-1} V]_{D^{(*)} K \to D_s^{(*)} \pi},
\label{Eq:BS-DK}\end{equation} 
where $V$ represents the $S$-wave driving potential, and $G$ is a diagonal matrix composed of two-meson scalar loop functions~\cite{Oller:2000fj,Oller:2000ma,Oller:1997ng}. In the numerical calculation, we employ the next-to-leading-order potential from Ref.~\cite{Altenbuchinger:2013vwa} (see Appendix \ref{DK-FSI}), with the pertinent low-energy constants and subtraction constant determined by fitting to the lattice QCD results of Ref.~\cite{Liu:2012zya}. For further details on the formulation of scattering between NGBs and heavy hadrons, we refer the reader to Refs.~\cite{Guo:2009ct,Guo:2015dha,Liu:2012zya,Altenbuchinger:2013vwa,Oller:2000fj,Yan:2018zdt}.

The rescattering amplitude for $T_{c\bar{s}1}^{*+} \to D_s^{*+} \pi^+ \pi^-$ via the $D_1^0 K^+ D^{*+}$-loop shown in Fig.~\ref{fig:triangle1}(b) shares a similar formulation to that of Fig.~\ref{fig:triangle1}(a), and is omitted here for brevity.

The rescattering processes shown in Figs.~\ref{fig:triangle1}(c) and (d) depict the contributions from $K\bar{K}\to \pi\pi$ final-state-interactions. The amplitude corresponding to the $D_1^+ K^0 \bar{K}^0$-loop in Fig.~\ref{fig:triangle1}(c) takes the form
\begin{align}
&\mathcal{A}^{[D_1^+ K^0 \bar{K}^0]}_{T_{c\bar{s}1}^{*+}  \to D_s^{*+} \pi^+ \pi^-}=-i\int\frac{d^4 q_1}{(2\pi)^4}\,\frac{\mathcal{A}(T_{c\bar{s}1}^{*+} \to D_1^+ K^0)}{(q_1^2 - m_{D_1}^2 + im_{D_1}\Gamma_{D_1})}
\nonumber \\[2mm]
& \times
\frac{\mathcal{A}(D_1^+ \to D_s^{*+} \bar{K}^0)}{(q_2^2 - m_{K}^2)}
\,
\frac{\mathcal{A}(K^0 \bar{K}^0\to \pi^+ \pi^-)}{(q_3^2 - m_{\bar{K}}^2)} \mathbb{F}(q_3^2)
,\label{Eq:loopKKpipi}
\end{align}
with the form factor $\mathbb{F}(q_3^2)=(m_{\bar{K}}^2-\Lambda^2)/(q_3^2-\Lambda^2)$.
The rescattering amplitude for Fig.~\ref{fig:triangle1}(d) shares a similar formalism.

The vertex function for $D_1^+ \to D_s^{*+} \bar{K}^0$ is also obtained within the framework of HHChPT,
\begin{align}
&\mathcal{A}(D_1^+ \to D_s^{*+} \bar{K}^0)
= \frac{4}{\sqrt{6}} \frac{h^\prime}{\Lambda_\chi  f_\pi} \sqrt{m_{D_1} m_{D_s^*}} \, 
 \nonumber\\
&\times \Big\{
3\, q_3^\lambda q_3^\nu+ \big[ (v \cdot q_3)^2 - q_3^2 \big] g^{\lambda \nu} 
\Big\} 
\, \varepsilon_\nu(q_1)\, \varepsilon_\lambda^*(p_2),
\end{align}
which is related to $\mathcal{A}({D_1 \to D^*\pi})$ via the SU(3) flavor symmetry.

The $K\bar{K} \to \pi\pi$ amplitude is unitarized using UChPT,
\begin{equation}
   \mathcal{A}(K^0 \bar{K}^0\to \pi^+ \pi^-) =[(1 - V G)^{-1} V]_{K^0 \bar{K}^0\to \pi^+ \pi^-}.
\end{equation}
We employ the same theoretical scheme and parameters as in Ref.~\cite{Liang:2014tia,Oller:1997ti}, focusing here on the $S$-wave coupled-channel interactions. The unitarized amplitude yields resonance poles, which may give rise to observable effects in the $\pi\pi$ and $D_s^*\pi$ spectra.

\subsubsection{$T_{c\bar{s}2}^{*+} \to D_s^{(*)+} \pi^+ \pi^-$}

Given that the $D_2$ meson decays into both $D^*\pi$ and $D\pi$ in a relative $D$-wave, the $T_{c\bar{s}2}^{*+}$ state is expected to decay into either $D_s^{*+} \pi^+ \pi^-$ or $D_s^{+} \pi^+ \pi^-$, as illustrated in Fig.~\ref{fig:triangle2} and Fig.~\ref{fig:triangle3}, respectively.

The rescattering amplitude of $T_{c\bar{s}2}^{*+} \to D_s^{*+} \pi^+ \pi^-$ via the $D_2^+K^0 D^{*0}$-loop in Fig.~\ref{fig:triangle2}(a) is given by 
\begin{align}
&\mathcal{A}^{[D_2^+ K^0 D^{*0}]}_{T_{c\bar{s}2}^{*+}  \to D_s^{*+} \pi^+ \pi^-}=-i\int\frac{d^4 q_1}{(2\pi)^4}\,\frac{\mathcal{A}(T_{c\bar{s}2}^{*+} \to D_2^+ K^0)}{(q_1^2 - m_{D_2}^2 + im_{D_2}\Gamma_{D_2})}
\nonumber \\[2mm]
& \times
\frac{\mathcal{A}(D_2^+ \to D^{*0} \pi^+)}{(q_2^2 - m_{K}^2)}
\,
\frac{\mathcal{A}(D^{*0} K^0 \to D_s^{*+} \pi^-)}{(q_3^2 - m_{D^*}^2)} \mathbb{F}(q_3^2)
. \label{Eq:loopT2Dsstar}
\end{align}
For the spin-2 state, the polarization sum is given by
\begin{align}
\sum_{\text{pol.}} \epsilon_{\mu\nu}\,\epsilon_{\alpha\beta}^{*}
=\frac{1}{2}\left(\tilde g_{\mu\alpha}\tilde g_{\nu\beta}
+\tilde g_{\mu\beta}\tilde g_{\nu\alpha}\right)
-\frac{1}{3}\tilde g_{\mu\nu}\tilde g_{\alpha\beta},
\end{align}
where $\tilde g_{\mu\nu}=-g_{\mu\nu}+v_\mu v_\nu$, and $v=(1,\boldsymbol{0})$ is adopted in the non-relativistic limit.
The amplitude of $D_2^+ K^0 \bar{K}^0$-loop in Fig.~\ref{fig:triangle2}(c) reads
\begin{align}
&\mathcal{A}^{[D_2^+ K^0 \bar{K}^0]}_{T_{c\bar{s}2}^{*+}  \to D_s^{*+} \pi^+ \pi^-}=-i\int\frac{d^4 q_1}{(2\pi)^4}\,\frac{\mathcal{A}(T_{c\bar{s}2}^{*+} \to D_2^+ K^0)}{(q_1^2 - m_{D_2}^2 + im_{D_2}\Gamma_{D_2})}
\nonumber \\[2mm]
& \times
\frac{\mathcal{A}(D_2^+ \to D_s^{*+} \bar{K}^0)}{(q_2^2 - m_{K}^2)}
\,
\frac{\mathcal{A}(K^0 \bar{K}^0\to \pi^+ \pi^-)}{(q_3^2 - m_{\bar{K}}^2)} \mathbb{F}(q_3^2)
. \label{Eq:loopT2KKpipi}
\end{align}
The amplitudes for the diagrams in Figs.~\ref{fig:triangle2}(b) and (d) take forms analogous to those shown in Eqs.~(\ref{Eq:loopT2Dsstar}) and ~(\ref{Eq:loopT2KKpipi}), respectively. The physical value $\Gamma_{D_2}=47.3$ MeV is adopted in the numerical calculation~\cite{ParticleDataGroup:2024cfk}.

Likewise, for the $T_{c\bar{s}2}^{*+} \to D_s^{+} \pi^+ \pi^-$ process, the rescattering amplitudes read
\begin{align}
&\mathcal{A}^{[D_2^+ K^0 D^{0}]}_{T_{c\bar{s}2}^{*+}  \to D_s^{+} \pi^+ \pi^-}=-i\int\frac{d^4 q_1}{(2\pi)^4}\,\frac{\mathcal{A}(T_{c\bar{s}2}^{*+} \to D_2^+ K^0)}{(q_1^2 - m_{D_2}^2 + im_{D_2}\Gamma_{D_2})}
\nonumber \\[2mm]
& \times
\frac{\mathcal{A}(D_2^+ \to D^{0} \pi^+)}{(q_2^2 - m_{K}^2)}
\,
\frac{\mathcal{A}(D^{0} K^0 \to D_s^{+} \pi^-)}{(q_3^2 - m_{D}^2)} \mathbb{F}(q_3^2), \label{Eq:loopT2Dspipi}
\end{align}
and
\begin{align}
&\mathcal{A}^{[D_2^+ K^0 \bar{K}^0]}_{T_{c\bar{s}2}^{*+}  \to D_s^{+} \pi^+ \pi^-}=-i\int\frac{d^4 q_1}{(2\pi)^4}\,\frac{\mathcal{A}(T_{c\bar{s}2}^{*+} \to D_2^+ K^0)}{(q_1^2 - m_{D_2}^2 + im_{D_2}\Gamma_{D_2})}
\nonumber \\[2mm]
& \times
\frac{\mathcal{A}(D_2^+ \to D_s^{+} \bar{K}^0)}{(q_2^2 - m_{K}^2)}
\,
\frac{\mathcal{A}(K^0 \bar{K}^0\to \pi^+ \pi^-)}{(q_3^2 - m_{\bar{K}}^2)} \mathbb{F}(q_3^2)
. \label{Eq:loopT2DsKKpipi}
\end{align}

The vertex functions for $D_2\to D^{(*)}\pi$ and $D_2\to D_s^{(*)}\bar{K}$ are obtained using the HHChPT,
\begin{align}
&\mathcal{A}(D_2^+ \to D^{*0} \pi^+) = - 4 i \frac{h^\prime}{\Lambda_\chi f_\pi} \sqrt{m_{D_2} m_{D^*}} \, 
 \nonumber\\
&\times \epsilon_{\lambda \nu \alpha \beta} v^{\alpha} p_1^{\lambda} p_{1 \mu} \varepsilon^{\mu \nu}(q_1) \, \varepsilon^{*\beta}(q_3), \\
&\mathcal{A}(D_2^+ \to D^{0} \pi^+) = 4 \frac{h^\prime}{\Lambda_\chi f_\pi} \sqrt{m_{D_2} m_{D}} \, 
 p_1^{\mu} p_1^{\nu} \varepsilon_{\mu \nu}(q_1), \\
&\mathcal{A}(D_2^+ \to D_s^{*+} \bar{K}^0) = - 4 i \frac{h^\prime}{\Lambda_\chi f_\pi} \sqrt{m_{D_2} m_{D_s^*}} \, 
 \nonumber\\
&\times \epsilon_{\lambda \nu \alpha \beta} v^{\alpha} q_3^{\lambda} q_{3 \mu} \varepsilon^{\mu \nu}(q_1) \, \varepsilon^{*\beta}(p_2), \\
&\mathcal{A}(D_2^+ \to D_s^{+} \bar{K}^0) = 4 \frac{h^\prime}{\Lambda_\chi f_\pi} \sqrt{m_{D_2} m_{D_s}} \, 
 q_3^{\mu} q_3^{\nu} \varepsilon_{\mu \nu}(q_1).
\end{align}

\subsection{Two-body decays of $T_{c\bar{s}1}^*$ and $T_{c\bar{s}2}^*$}

In addition to three-body decays, isospin-violating two-body decays offer a sensitive probe of molecular structure, especially when the parent state lies close to threshold and couples strongly to its constituents~\cite{Guo:2017jvc}. The decay $T_{c\bar{s}1}^* \to D_{s1}(2460)\,\pi^0$ is particularly instructive: if $T_{c\bar{s}1}^*$ and $D_{s1}(2460)$ are dominated by $D_1K$ and $D^*K$ components, respectively, the process represents a transition from one hadronic molecule to another. In the isospin limit, the two contributing loop amplitudes cancel exactly, and the residual amplitude arises solely from small mass splittings between charged and neutral intermediate states. The absolute rates and their pattern across related channels—such as $T_{c\bar{s}2}^* \to D_{s1}(2460)\,\pi^0$ and $T_{c\bar{s}2}^* \to D_{s0}^*(2317)\,\pi^0$—thus provide clean complementary probes of long-range molecular dynamics.

\begin{figure}[htbp]
\centering
\includegraphics[width=1.\hsize]{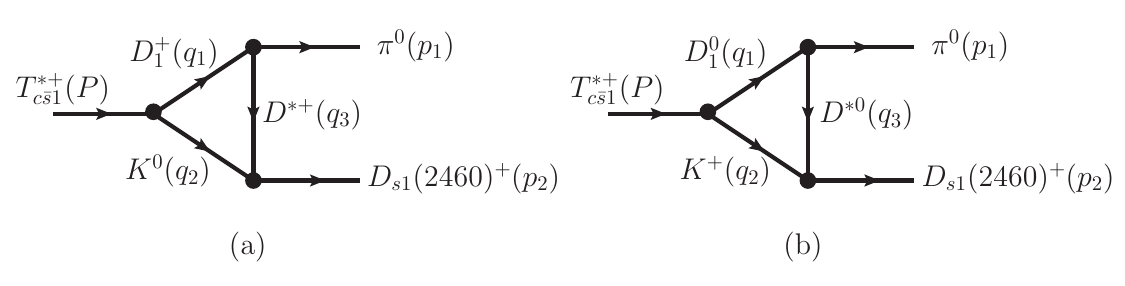}
\caption{Triangle diagrams for $T_{c\bar{s}1}^{*+} \to D_{s1}(2460)^+ \pi^0$: (a) $D_1^+ K^0 D^{*+}$ loop; (b) $D_1^0 K^+ D^{*0}$ loop.}
\label{fig:twobody-Tcs1}
\end{figure}

The interference between the two diagrams in Fig.~\ref{fig:twobody-Tcs1} is destructive. Since $D_{s1}(2460)$ is an isoscalar state, its coupling to the $D^*K$ channel is described by the Lagrangian
\begin{align}
\mathcal{L}_{D_{s1} D^* K} &= \frac{g_{D_{s1}}} {\sqrt{2}}\, \bar{D}_{s1}^{\mu} \left( D_{\mu} ^{*+} K^0 + D_{\mu} ^{*0} K^+ \right) + \text{H.c.} \label{Eq:lagDs12460}
\end{align}
The corresponding amplitudes for Figs.~\ref{fig:twobody-Tcs1}(a) and (b) read
\begin{align}
&\mathcal{A}_{T_{c\bar{s}1}^{*+}\to D_{s1}^+ \pi^0}^{[D_1^+ K^0 D^{*+}]} = -i \int \frac{d^4 q}{(2\pi)^4} \frac{\mathcal{A}(T_{c\bar{s}1}^{*+} \to D_1^+ K^0)}{(q_1^2 - m_{D_1^+}^2)} \nonumber \\[2mm]
& \times \frac{\mathcal{A}(D_1^+ \to D^{*+} \pi^0)}{(q_2^2 - m_{K^0}^2)} \frac{\mathcal{A}(D^{*+} K^0 \to D_{s1}^+)}{(q_3^2 - m_{D^{*+}}^2)} \, \mathbb{F}(q_3^2), \\[4mm]
&\mathcal{A}_{T_{c\bar{s}1}^{*+}\to D_{s1}^+ \pi^0}^{[D_1^0 K^+ D^{*0}]} = -i \int \frac{d^4 q}{(2\pi)^4} \frac{\mathcal{A}(T_{c\bar{s}1}^{*+} \to D_1^0 K^+)}{(q_1^2 - m_{D_1^0}^2)} \nonumber \\[2mm]
& \times \frac{\mathcal{A}(D_1^0 \to D^{*0} \pi^0)}{(q_2^2 - m_{K^{+}}^2)} \frac{\mathcal{A}(D^{*0} K^+ \to D_{s1}^+)}{(q_3^2 - m_{D^{*0}}^2)} \, \mathbb{F}(q_3^2),
\end{align}
with $\mathbb{F}(q_3^2)=(m_{D^*}^2-\Lambda^2)/(q_3^2-\Lambda^2)$. In the isospin-symmetry limit, the two amplitudes cancel exactly; the small mass differences among intermediate states lead to incomplete cancellation and thus isospin violation.

\begin{figure}[htbp]
\centering
\includegraphics[width=1.0\hsize]{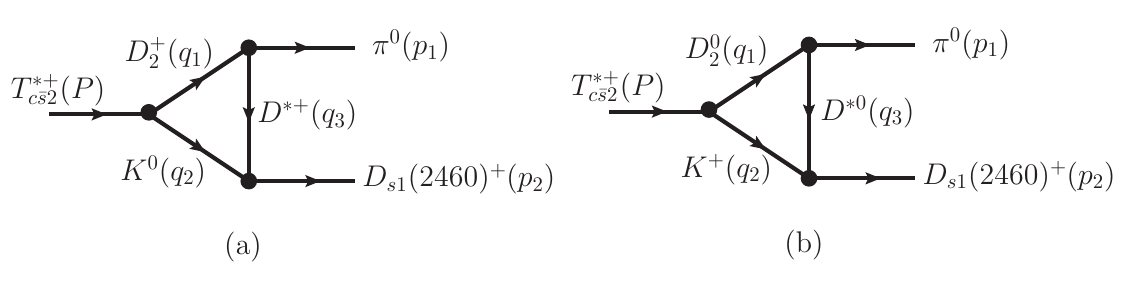}
\caption{Triangle diagrams for $T_{c\bar{s}2}^{*+} \to D_{s1}(2460)^+ \pi^0$: (a) $D_2^+ K^0 D^{*+}$ loop; (b) $D_2^0 K^+ D^{*0}$ loop.}
\label{fig:twobody-Tcs2-Ds1}
\end{figure}

\begin{figure}[htbp]
\centering
\includegraphics[width=1.0\hsize]{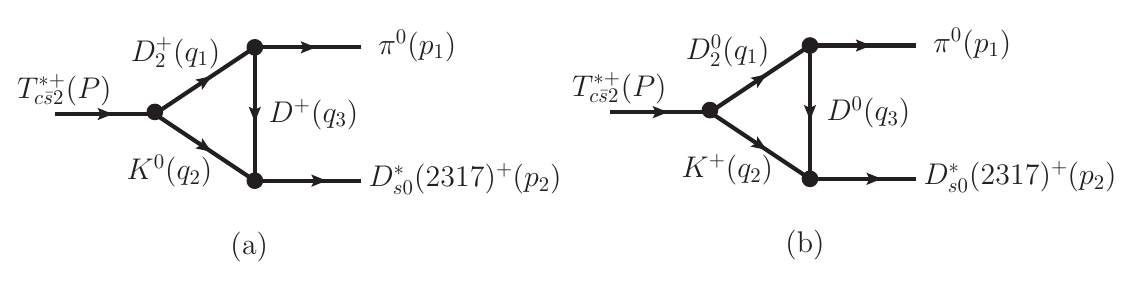}
\caption{Triangle diagrams for $T_{c\bar{s}2}^{*+} \to D_{s0}^{*}(2317)^+ \pi^0$: (a) $D_2^+ K^0 D^{+}$ loop; (b) $D_2^0 K^+ D^{0}$ loop.}
\label{fig:twobody-Tcs2-Ds0}
\end{figure}

The isospin-violating decays of $T_{c\bar{s}2}^{*+}$ follow a similar pattern. The key difference is that $T_{c\bar{s}2}^{*+}$ can decay into both $D_{s1}(2460)^+\pi^0$ and $D_{s0}^*(2317)^+\pi^0$, as illustrated in Figs.~\ref{fig:twobody-Tcs2-Ds1} and \ref{fig:twobody-Tcs2-Ds0}, respectively. The coupling of $D_{s0}^*(2317)$ to the $DK$ channel is described by
\begin{align}
\mathcal{L}_{D_{s0}^* D K} &= \frac{g_{D_{s0}^*}} {\sqrt{2}}\, \bar{D}_{s0}^* \left( D^{+} K^0 + D^{0} K^+ \right) + \text{H.c.} \label{Eq:lagDs02317}
\end{align}

\section{Numerical Results}

\begin{figure}[htbp]
\centering
\includegraphics[width=0.85\hsize]{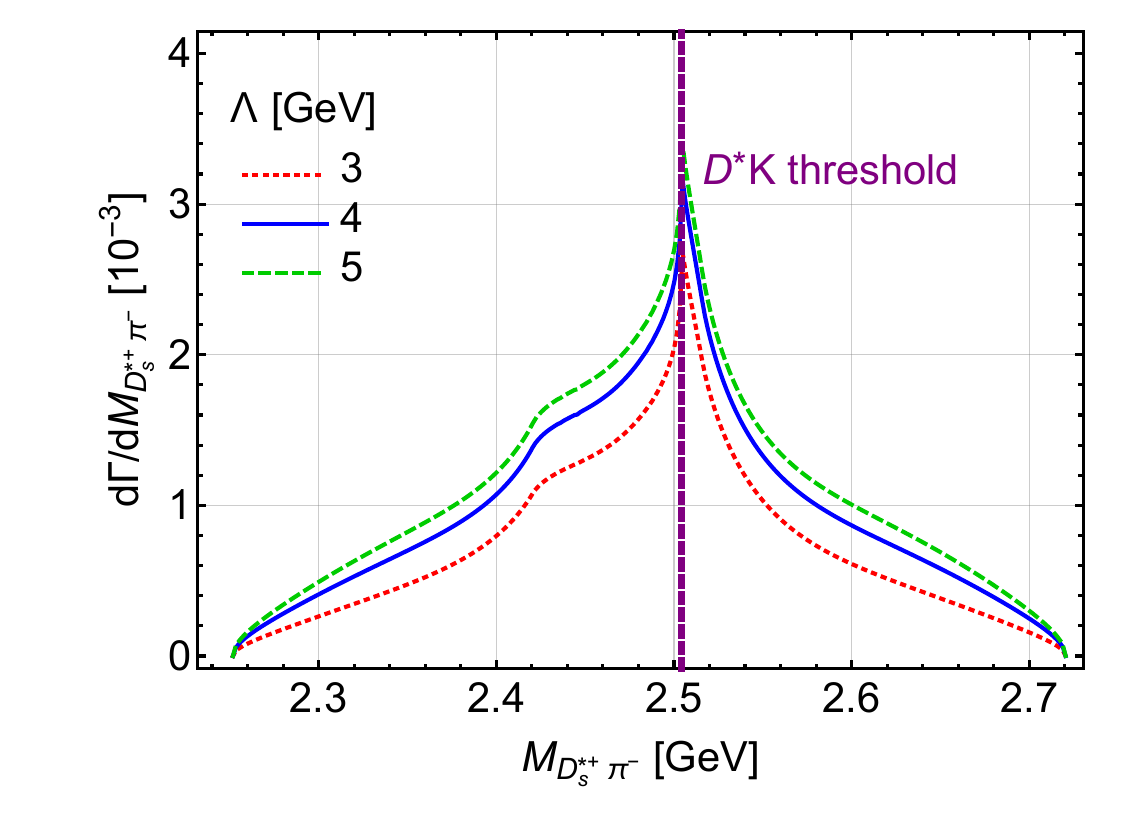}
\caption{The $\Lambda$ dependence of the $D_s^{*+}\pi^-$ invariant mass distribution for the decay $T_{c\bar{s}1}^{*+} \to D_s^{*+}\pi^+\pi^-$. }
\label{fig:Lambda}
\end{figure}

\begin{figure}[htbp]
\centering
\includegraphics[width=0.85\hsize]{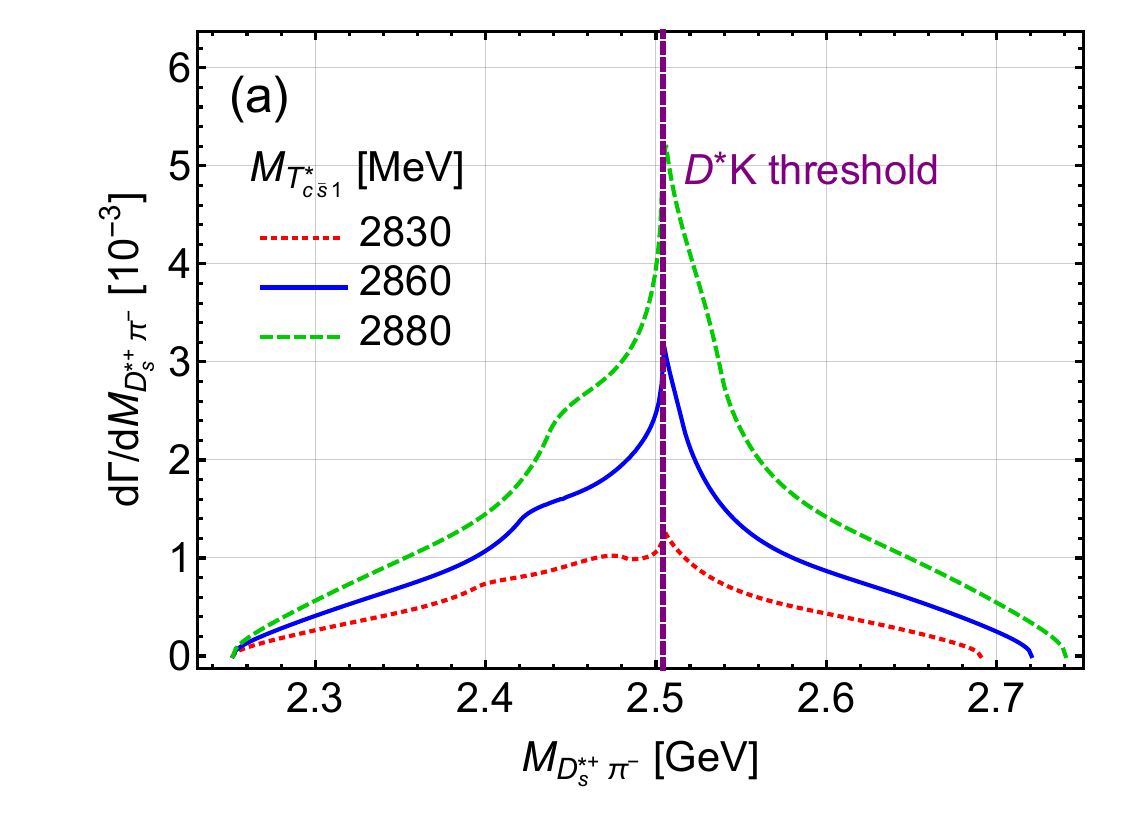}
\includegraphics[width=0.85\hsize]{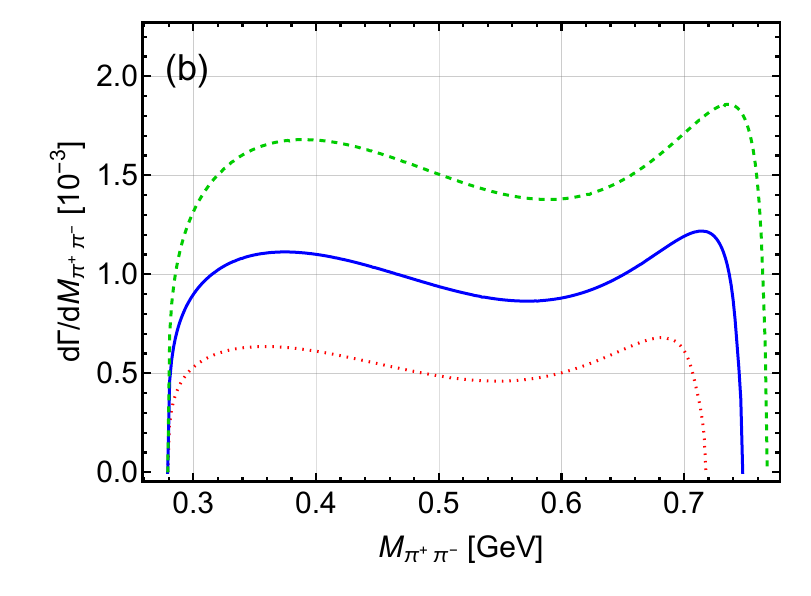}
\caption{Invariant mass distributions of (a) $D_s^{*+}\pi^-$ and (b) $\pi^+\pi^-$ for the decay $T_{c\bar{s}1}^{*+} \to D_s^{*+}\pi^+\pi^-$, computed for different masses of the $T_{c\bar{s}1}^{*+}$ state. The cutoff is fixed at $\Lambda = 4$ GeV.}
\label{fig:massDist}
\end{figure}

\begin{figure}[htbp]
\centering
\includegraphics[width=0.85\hsize]{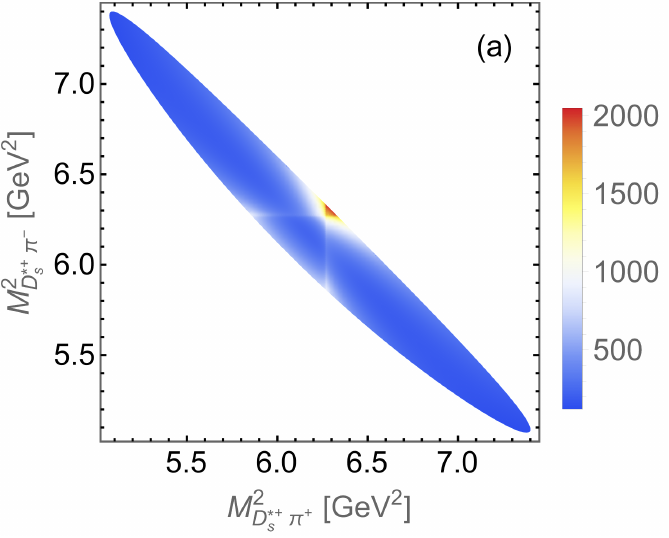}
\includegraphics[width=0.85\hsize]{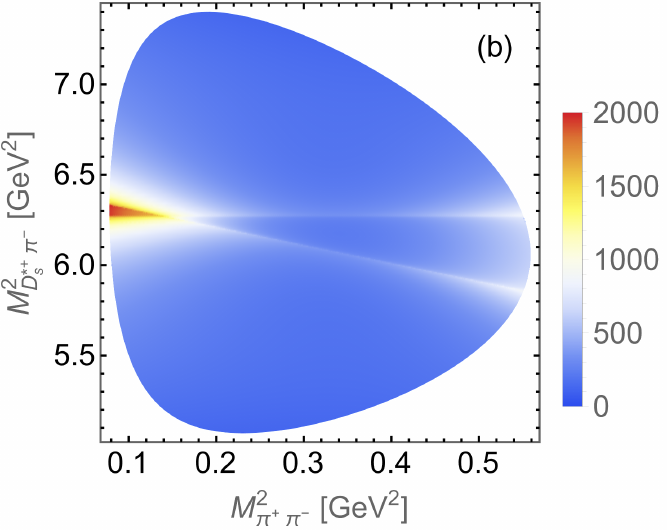}
\caption{Dalitz plots for the decay $T_{c\bar{s}1}^{*+} \to D_s^{*+}\pi^+\pi^-$: (a) $M_{D_s^{*+}\pi^-}^2$ versus $M_{D_s^{*+}\pi^+}^2$ and (b) $M_{D_s^{*+}\pi^-}^2$ versus $M_{\pi^+\pi^-}^2$,  with the mass of $T_{c\bar{s}1}^{*+}$ fixed at $M=2860~\mathrm{MeV}$ and the cutoff parameter fixed at $\Lambda = 4~\mathrm{GeV}$.}
\label{fig:T1Dalitz}
\end{figure}

\begin{figure}[tb]
	\centering
\includegraphics[width=0.8\linewidth]{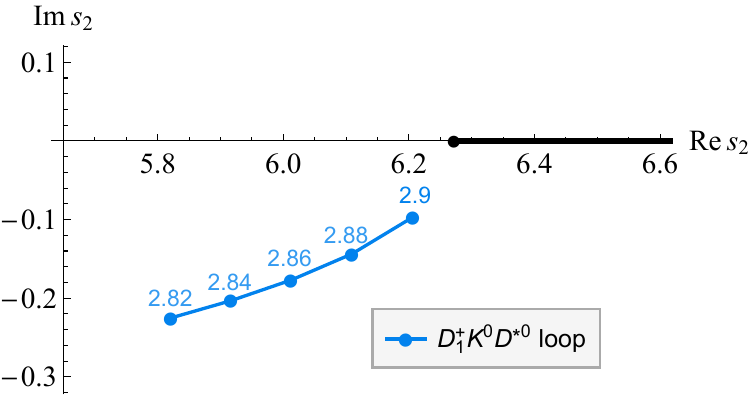}
	\caption{Trajectory of the TS location corresponding to Fig.~\ref{fig:triangle1}(a) in the complex $s_2$ plane, as a function of the $T_{c\bar{s}1}^*$ mass varied from $2.8$ to $2.9$~GeV, with $s_2\equiv (p_2+p_3)^2$. The thick black line along the real axis denotes the unitarity cut starting at the $D^*K$ threshold. Annotations indicate the corresponding $T_{c\bar{s}1}^*$ mass values in GeV.}
	\label{fig:TS_traj}
\end{figure}

\begin{figure}[htbp]
\centering
\includegraphics[width=0.85\hsize]{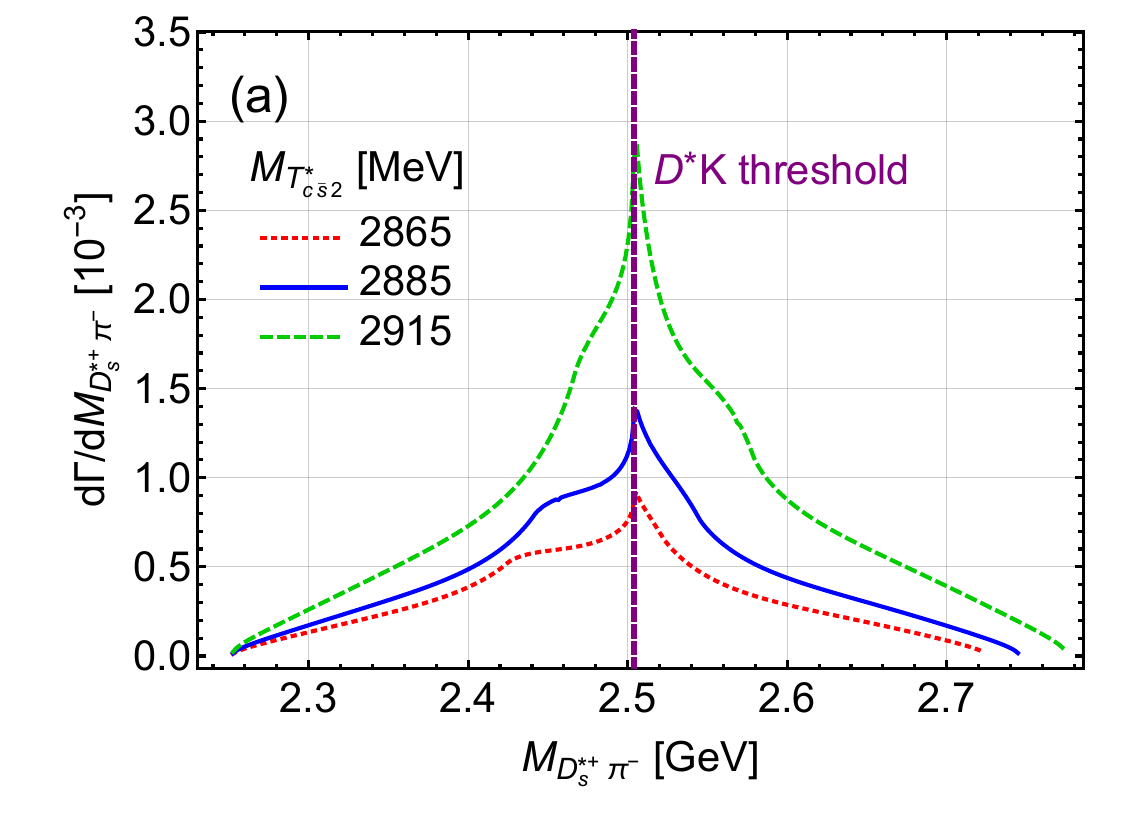}
\includegraphics[width=0.85\hsize]{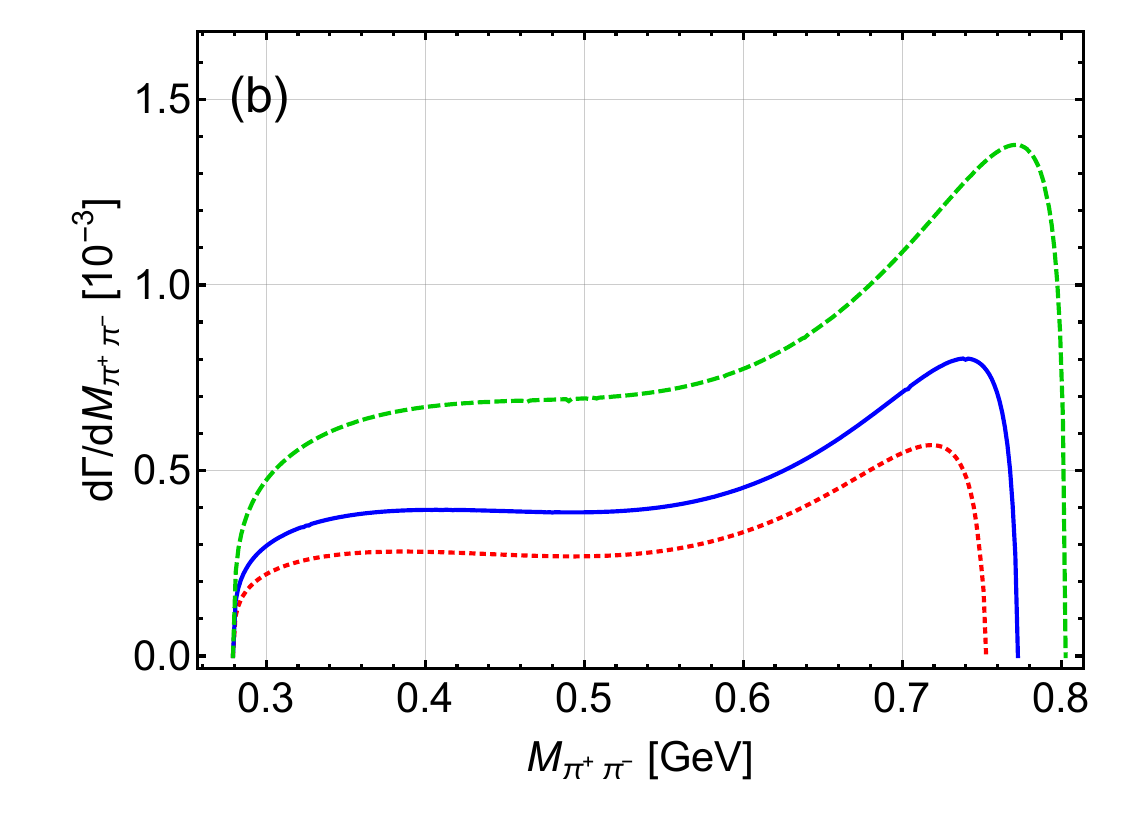}
\caption{Invariant mass distributions of (a) $D_s^{*+}\pi^-$ and (b) $\pi^+\pi^-$ for the decay $T_{c\bar{s}2}^{*+} \to D_s^{*+}\pi^+\pi^-$, computed for different masses of the $T_{c\bar{s}2}^{*+}$ state. The cutoff is fixed at $\Lambda = 4$ GeV.}
\label{fig:Tcs2_Dsstar}
\end{figure}

\begin{figure}[htbp]
\centering
\includegraphics[width=0.85\hsize]{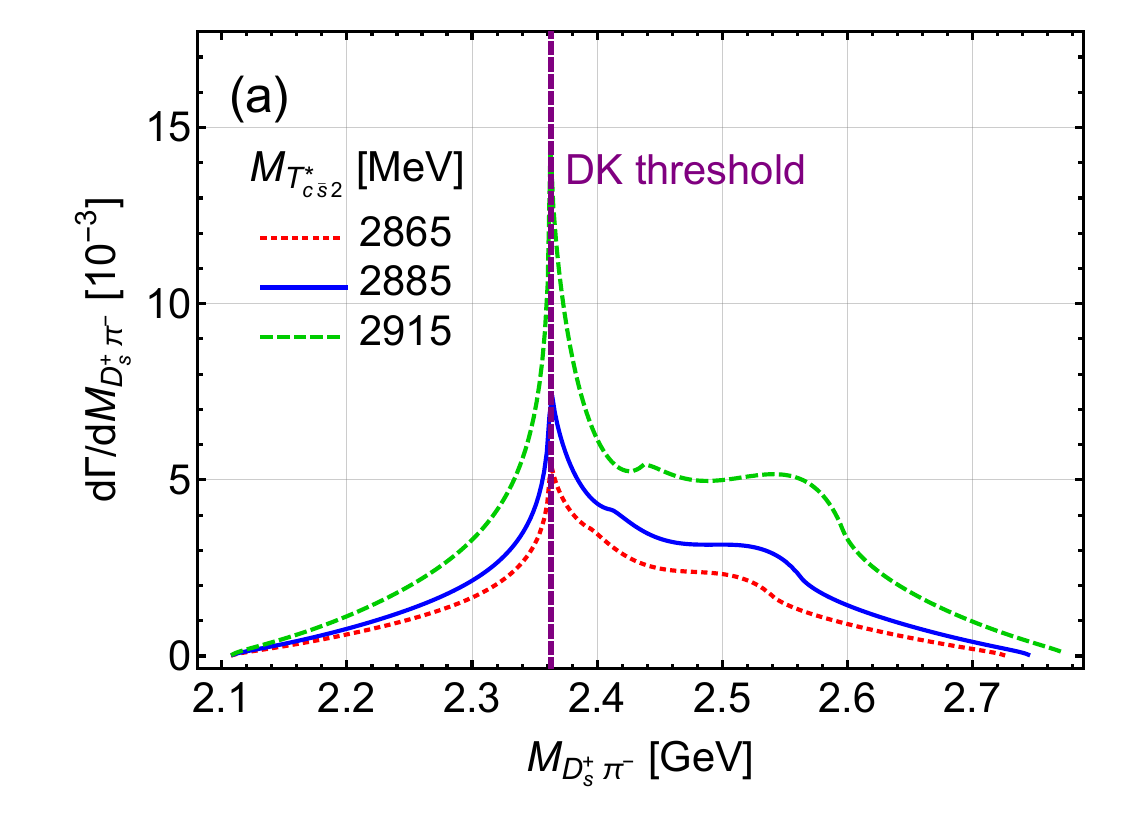}
\includegraphics[width=0.85\hsize]{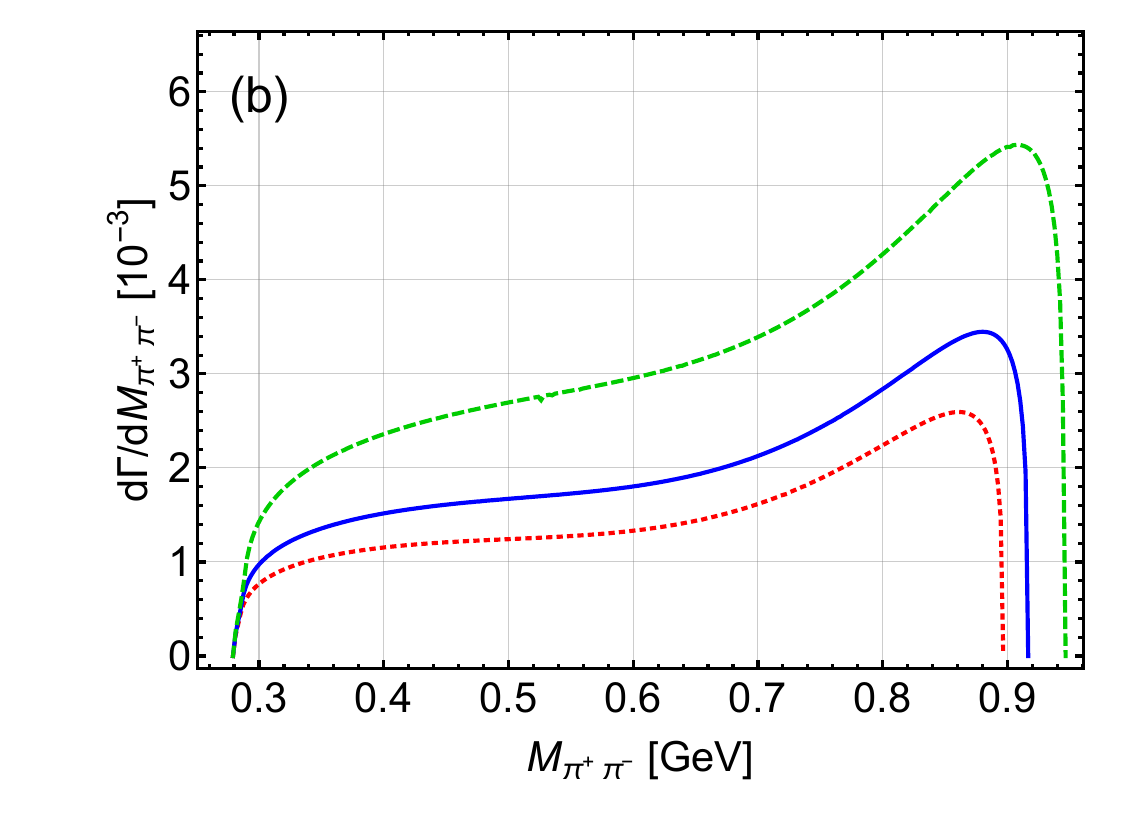}
\caption{Invariant mass distributions of (a) $D_s^{+}\pi^-$ and (b) $\pi^+\pi^-$ for the decay $T_{c\bar{s}2}^{*+} \to D_s^{+}\pi^+\pi^-$, computed for different masses of the $T_{c\bar{s}2}^{*+}$ state. The cutoff is fixed at $\Lambda = 4$ GeV.}
\label{fig:Tcs2_Ds}
\end{figure}

\subsection{ Resonance-like structures in three-body decays}

The molecular couplings $g_{T_1}$, $g_{T_2}$, $g_{D_{s0}^*}$ and $g_{D_{s1}}$ are obtained from Ref.~\cite{Wang:2025jcq}, as shown in Table~\ref{tab:coupling}.  There is still a free parameter, i.e., the cutoff energy $\Lambda$, in the form factors of the rescattering amplitudes. Its explicit value should be determined from the experimental data.
Figure~\ref{fig:Lambda} illustrates the dependence of the $D_s^{*+}\pi^-$ invariant mass distribution on the cutoff parameter $\Lambda$ in the form factor. The value of $\Lambda$ is typically chosen to be larger than $m_{\text{ex}}$, the mass of the corresponding exchanged particle. For the diagrams in Figs.~\ref{fig:triangle1}(a) and (b), however, adopting a smaller $\Lambda$ (while still keeping $\Lambda > m_{\text{ex}}$) would introduce unphysical kinematic singularities that strongly affect the physical amplitudes. Therefore, we employ a relatively larger cutoff. As $\Lambda$ increases from $3$~GeV to $5$~GeV, both the line shape and the magnitude of the differential width exhibit only mild variations. This indicates that the numerical results are not strongly sensitive to the detailed modeling of off-shell effects. The stability can be attributed to the fact that in the dominant triangle diagrams of Figs.~\ref{fig:triangle1}(a) and (b), the three internal particles can become simultaneously on-shell, thereby suppressing the sensitivity to the form factor. In contrast, the contributions from the $D_1K\bar{K}$ loops shown in Figs.~\ref{fig:triangle1}(c) and (d) are found to be significantly weaker, since the intermediate states in these diagrams cannot be simultaneously on-shell under the same kinematic conditions. The cutoff $\Lambda$ is fixed at 4 GeV in the following discussions.

\begin{table}[h]
\centering
\caption{Coupling constants of the molecular states to their components, in units of GeV.}
\label{tab:coupling}
\begin{tabular}{cccc}
\hline\hline
$g_{T1}$ & $g_{T2}$ & $g_{D_{s0}^*}$ & $g_{D_{s1}}$  \\
14 & 14 & 11 & 12  \\
\hline\hline
\end{tabular}
\end{table}

Figure~\ref{fig:massDist} displays the $D_s^{*}\pi$ and $\pi\pi$ invariant mass distributions for the decay $T_{c\bar{s}1}^{*}\to D_s^{*}\pi\pi$, computed for several masses of the $T_{c\bar{s}1}^{*}$ state. In the $D_s^{*}\pi$ spectrum, a pronounced and narrow peak appears near the $D^{*}K$ threshold. This structure originates from the TS arising in the rescattering amplitudes of the $D_1KD^{*}$ loop diagrams. The bump observed around $2.42$~GeV is a reflection effect from the Dalitz plot kinematics. In the $\pi\pi$ invariant mass distribution, two bumps emerge at the low- and high-mass ends. These also arise from kinematical reflections, corresponding to the resonance-like structures seen in the $D_s^{*+}\pi^+$ and $D_s^{*+}\pi^-$ distributions, as shown in the Dalitz plot Fig.~\ref{fig:T1Dalitz}.

The narrow peak observed in the $D_s^{*}\pi$ invariant mass distribution, as shown in Fig.~\ref{fig:massDist}(a), is interpreted as a manifestation of the TS rather than a genuine resonance. This conclusion is supported by the following considerations. First, the interaction in the $(S,I)=(1,1)$ channel is relatively weak. Although some theoretical studies suggest the possible existence of resonances in this channel~\cite{Guo:2009ct,Guo:2015dha}, any such states are expected to have large widths and thus would not produce the narrow structure seen in our calculation. Second, the kinematic conditions for the TS can be examined through the trajectory of the singularity pole. As illustrated in Fig.~\ref{fig:TS_traj}, as the mass of the $T_{c\bar{s}1}^{*}$ state varies within the range expected for a $D_1K$ molecule, the singularity does not lie exactly on the real axis but remains sufficiently close to the physical region. This proximity leads to a significant enhancement of the rescattering amplitude around the $D^*K$ threshold, giving rise to a pronounced narrow structure in the invariant mass distribution. Consequently, the predicted narrow peak in the $D_s^{*}\pi$ spectrum can serve as a distinctive signature for identifying $D_1K$ molecular states in future experiments.

Similar resonance-like features appear in the decays of the $T_{c\bar{s}2}^*$ state. As shown in Figs.~\ref{fig:Tcs2_Dsstar} and~\ref{fig:Tcs2_Ds}, the $D_s^{*}\pi$ and $D_s\pi$ invariant mass distributions for $T_{c\bar{s}2}^{*+} \to D_s^{*+}\pi^+\pi^-$ and $T_{c\bar{s}2}^{*+} \to D_s^{+}\pi^+\pi^-$, respectively, exhibit narrow peaks analogous to those observed in the $T_{c\bar{s}1}^*$ case. In the $D_s^{*}\pi$ spectrum, the peak is located near the $D^*K$ threshold, while in the $D_s\pi$ spectrum, the peak appears near the $DK$ threshold. Both structures originate from TSs in the $D_2KD^{*}$ and $D_2KD$ loop diagrams, respectively. As in the $T_{c\bar{s}1}^*$ decay, the $\pi\pi$ invariant mass distributions for these $T_{c\bar{s}2}^*$ decays also display two bumps at the low- and high-mass ends, which arise from kinematical reflections in the Dalitz plot, corresponding to the resonance-like structures in the $D_s^{(*)}\pi^\pm$ combinations. These features further confirm that the observed peaks are not genuine resonances but rather manifestations of TSs. The similarity between the $T_{c\bar{s}1}^*$ and $T_{c\bar{s}2}^*$ decays reinforces the interpretation of both states as hadronic molecules of $D_1K$ and $D_2K$, respectively.

\subsection{Isospin violation decays}

\begin{figure*}[htbp]
\centering
\includegraphics[width=0.32\textwidth]{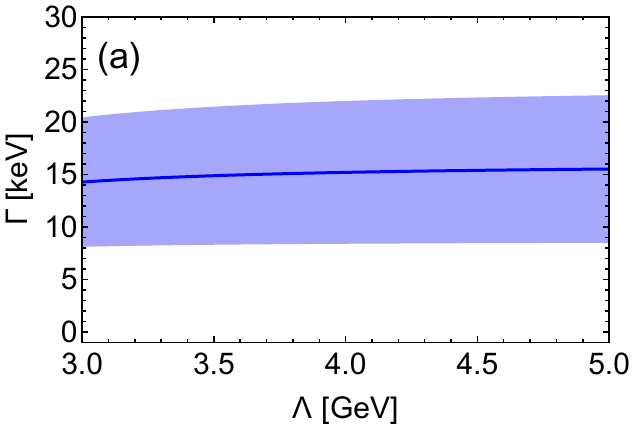}\hfill
\includegraphics[width=0.32\textwidth]{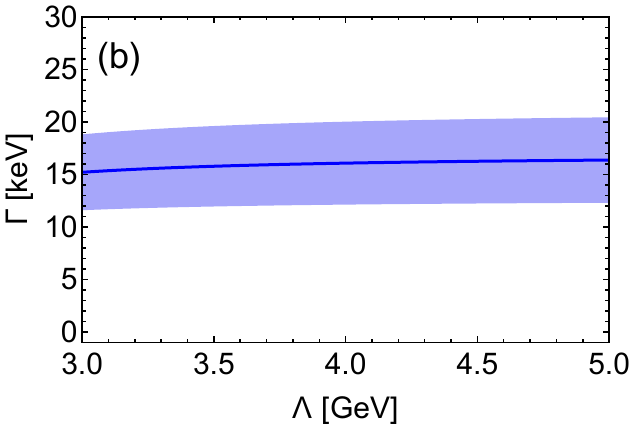}\hfill
\includegraphics[width=0.32\textwidth]{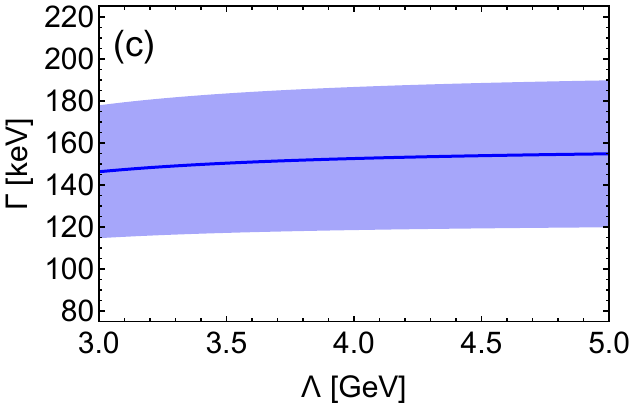}
\caption{Partial decay widths of (a) $T_{c\bar{s}1}^{*+} \to D_{s1}(2460)^+\pi^0$, (b) $T_{c\bar{s}2}^{*+} \to D_{s1}(2460)^+\pi^0$, and (c) $T_{c\bar{s}2}^{*+} \to D_{s0}^*(2317)^+\pi^0$ as functions of the cutoff parameter $\Lambda$. The solid lines correspond to the central values of the input masses from Ref.~\cite{ParticleDataGroup:2024cfk}, while the shaded bands account for the uncertainties induced by the mass variations of the charged and neutral intermediate mesons: $D_1(2420)$ in panel (a), and $D_2^*(2460)$ in panels (b) and (c).}
\label{fig:partialwidthLambda}
\end{figure*}

Figure~\ref{fig:partialwidthLambda}(a) shows the partial decay width of $T_{c\bar{s}1}^{*+} \to D_{s1}(2460)^+\pi^0$ as a function of the cutoff parameter $\Lambda$ in the form factor. With the mass of $T_{c\bar{s}1}^{*+}$ fixed at 2860~MeV, the width grows moderately as $\Lambda$ increases from $3$~GeV to $5$~GeV, staying around $15$~keV. This mild dependence on $\Lambda$ indicates that the result is robust against the detailed modeling of off-shell effects.

Similar features appear in the decays of the $T_{c\bar{s}2}^*$ state. The processes $T_{c\bar{s}2}^{*+} \to D_{s1}(2460)^+ \pi^0$ and $T_{c\bar{s}2}^{*+} \to D_{s0}^*(2317)^+ \pi^0$ also proceed via isospin-violating triangle loops, with the latter involving $DK$ rescattering to the $D_{s0}^*(2317)$ molecule. As illustrated in Figs.~\ref{fig:partialwidthLambda}(b) and~\ref{fig:partialwidthLambda}(c), with the mass of $T_{c\bar{s}2}^{*+}$ fixed at 2915~MeV, the partial widths for these two channels are approximately $16$~keV and $150$~keV, respectively.

For these isospin-violating processes, the obtained partial widths range from about $15$~keV to $150$~keV, which is considerably larger than typical isospin-violating decay widths. This enhancement is even more striking when one considers that, in the heavy quark limit, the $D_1 \to D^*\pi$, $D_2 \to D^*\pi$, and $D_2 \to D\pi$ decays are all $D$-wave, as required by angular momentum conservation and heavy quark symmetry. Consequently, the three isospin-violating decays discussed above proceed via $D$-wave, which normally leads to additional suppression. The fact that we still obtain such sizable partial widths despite this $D$-wave suppression indicates remarkably strong couplings, reflecting the molecular nature of the initial and final states. Thus, these relatively large isospin-violating partial widths can serve as a characteristic signature of hadronic molecule decays.

The isospin-violating two-body decays provide a clean discriminant between molecular and compact interpretations: for a molecular state, the partial widths are enhanced by the strong coupling to its constituent hadrons, whereas for a conventional $c\bar{s}$ state, no such enhancement is expected.

\section{Summary}

We have systematically investigated the two- and three-body decays of the $D_1K$ and $D_2K$ molecular states with isospin $I=0$, denoted as $T_{c\bar{s}1}^*$ and $T_{c\bar{s}2}^*$, into final states containing $D_s^{(*)}$ mesons and pions. Our calculations are based on the HHChPT for the vertices and the UChPT for the $S$-wave $D^{(*)}K\to D_s^{(*)}\pi$ rescattering amplitudes.

For $T_{c\bar{s}1}^*$, the three-body decay $T_{c\bar{s}1}^*\to D_s^*\pi\pi$ proceeds via the $D_1KD^*$ triangle loop and produces a narrow TS peak near the $D^*K$ threshold in the $D_s^*\pi$ spectrum, while the $\pi\pi$ spectrum shows reflection-induced bumps. Similar TS peaks appear in $T_{c\bar{s}2}^*$ decays: $T_{c\bar{s}2}^*\to D_s^*\pi\pi$ and $T_{c\bar{s}2}^*\to D_s\pi\pi$ peak near the $D^*K$ and $DK$ thresholds, respectively.

Isospin-violating two-body decays provide a clean molecular signature. The partial widths for $T_{c\bar{s}1}^*\to D_{s1}(2460)\pi^0$, $T_{c\bar{s}2}^*\to D_{s1}(2460)\pi^0$, and $T_{c\bar{s}2}^*\to D_{s0}^*(2317)\pi^0$ are about $15$~keV, $16$~keV, and $150$~keV, respectively. These values are considerably larger than typical isospin-violating widths, especially given the $D$-wave suppression of the involved vertices, reflecting the strong couplings inherent to molecular states.

Our results also highlight the close connection between the $D_1K$/$D_2K$ molecules and the well-established $D_{s0}^*(2317)$ and $D_{s1}(2460)$ states, which are widely interpreted as $DK$ and $D^*K$ molecules. The decay patterns of $T_{c\bar{s}1}^*$ and $T_{c\bar{s}2}^*$ mirror those of the lighter molecular states, providing a unified picture of heavy-light meson dynamics governed by chiral and heavy-quark symmetries.

The narrow TS-induced peaks in the three-body invariant mass distributions and the relatively large isospin-violating two-body decay widths provide complementary and clean probes for identifying $D_1K$ and $D_2K$ molecular states. These predictions can be tested in future high-statistics experiments at LHCb, Belle II, and BESIII, and will help distinguish molecular configurations from conventional $c\bar{s}$ charmed-strange mesons or other exotic interpretations.

\begin{acknowledgments}
This work is supported by the National Natural Science Foundation of China under Grants  No.~12235018 and No.~11975165.
\end{acknowledgments}

\begin{appendix}
\begin{widetext}

\section{$D^{(*)} K \to D_s^{(*)} \pi$ interactions}
\label{DK-FSI}

Following Refs.~\cite{Liu:2012zya,Altenbuchinger:2013vwa}, we employ the effective chiral Lagrangian to describe the interaction of charmed mesons with NGBs. In this work, we work with the driving potential $V$ in Eq.~(\ref{Eq:BS-DK}) expanded to next-to-leading order (NLO). 
Explicitly, the leading-order (LO) and NLO contributions to the potential are given by
\begin{equation}
\begin{aligned}
&V_{\mathrm{LO}}\big(P(p_1)\phi(p_2)\rightarrow P(p_3)\phi(p_4)\big)=\frac{1}{4F_0^2}\,\mathcal{C}_{\mathrm{LO}}\,(s-u),\\[4pt]
&V_{\mathrm{NLO}}\big(P(p_1)\phi(p_2)\rightarrow P(p_3)\phi(p_4)\big) = \\
&-\frac{8}{F_0^2}\,\mathcal{C}_{24}\bigg[ c_2\,p_2\!\cdot\!p_4 - \frac{c_4}{m_P^2}\big(p_1\!\cdot\!p_4\;p_2\!\cdot\!p_3 + p_1\!\cdot\!p_2\;p_3\!\cdot\!p_4\big) \bigg] \\
& -\frac{4}{F_0^2}\,\mathcal{C}_{35}\bigg[ c_3\,p_2\!\cdot\!p_4 - \frac{c_5}{m_P^2}\big(p_1\!\cdot\!p_4\;p_2\!\cdot\!p_3 + p_1\!\cdot\!p_2\;p_3\!\cdot\!p_4\big) \bigg] \\
& -\frac{8}{F_0^2}\,\mathcal{C}_0\,c_0 + \frac{4}{F_0^2}\,\mathcal{C}_1\,c_1,
\end{aligned}
\end{equation}
where $s=(p_1+p_2)^2=(p_3+p_4)^2$ and $u=(p_1-p_4)^2=(p_3-p_2)^2$. The coefficients $\mathcal{C}_i$ for the $(S,I)=(1,1)$ channel and the low-energy constants $c_i$ are listed in Table~\ref{table:C} and Table~\ref{table:c}, respectively. The decay constant is set to $F_0 = 92.2$ MeV.

A modified loop function $G_{\text{HQS}}$, which ensures heavy-quark symmetry (HQS) and consistent chiral power counting, is adopted and reads
\begin{equation}
\begin{aligned}
G_{\mathrm{HQS}}(s,M^2,m^2)&=G_{\overline{\mathrm{MS}}}(s,M^2,m^2) -\frac{1}{16\pi^2}
\Bigg(\!\log\frac{\tilde{M}^2}{\mu^2}-2\Bigg)\\
&\quad+\frac{m_{\mathrm{sub}}}{16\pi^2\tilde{M}}
\Bigg(\!\log\frac{\tilde{M}^2}{\mu^2}+a\Bigg),\\[6pt]
G_{\overline{\mathrm{MS}}}(s,M^2,m^2) &= \frac{1}{16\pi^2}\Bigg\{
\frac{m^2-M^2+s}{2s}\log\frac{m^2}{M^2}-\frac{q}{\sqrt{s}}\\
&\quad\Big( \log\big[2q\sqrt{s}+m^2-M^2-s\big]
+\log\big[2q\sqrt{s}-m^2+M^2-s\big] \\
&\quad -\log\big[2q\sqrt{s}+m^2-M^2+s\big]
-\log\big[2q\sqrt{s}-m^2+M^2+s\big]\Big) \\
&\quad +\Big(\log\frac{M^2}{\mu^2}-2\Big) \Bigg\}.
\end{aligned}
\end{equation}
Here, $q = \sqrt{\big[s-(M+m)^2\big]\big[s-(M-m)^2\big]}\big/(2\sqrt{s})$, with $M$ and $m$ denoting the masses of the charmed meson and the NGB in the intermediate state, respectively. The renormalization scale is fixed at $\mu = 1$ GeV, and the subtraction constant is taken as $a = -4.13$. The mass scale $m_{\mathrm{sub}}$ is set to the $\rm{SU(3)}$-averaged NGB mass, defined by $m_{\mathrm{sub}} = (3m_{\pi}+4m_{K}+m_{\eta})/8 \approx 370.4$ MeV.

\begin{table}[h]
\centering
\caption{Coefficients of the LO and NLO potentials for the $(S, I) = (1, 1)$ channel.}
\begin{tabular}{lcccccc}
\hline\hline
(S, I) & Channel & $\mathcal{C}_{LO}$ & $\mathcal{C}_{0}$ & $\mathcal{C}_{1}$ & $\mathcal{C}_{24}$ & $\mathcal{C}_{35}$ \\
\hline
(1, 1) & $DK \to DK$ & $m_K^2$ & 0 & 0 & 1 & 0\\
       & $D_s \pi \to D_s \pi$ & $m_\pi^2$ & 0 & 0 & 1 & 0 \\
       & $DK \to D_s \pi$ & 0 & 1 & $\frac{1}{2}(m_K^2 + m_\pi^2)$ & 0 & $-1$ \\
\hline\hline
\end{tabular}
\label{table:C}
\end{table}

\begin{table}[h]
\centering
\caption{Central values of the low-energy constants $c_i$~\cite{Altenbuchinger:2013vwa}.}
\label{tab:lecs_central_values}
\begin{tabular}{cccccc}
\hline\hline
$c_0$ & $c_1$ & $c_2$ & $c_3$ & $c_4$ & $c_5$ \\
0.015 & $-0.214$ & 0.036 & $-1.931$ & 0.052 & $-0.96$ \\
\hline\hline
\end{tabular}
\label{table:c}
\end{table}

\end{widetext}
\end{appendix}

\end{document}